\documentclass[reprint, superscriptaddress, eprint, twocolumn, amsmath, amssymb, prx]{revtex4-2}
\usepackage[english]{babel}
\usepackage{graphicx}
\usepackage{dcolumn}
\usepackage{bm}
\usepackage{dsfont}

\usepackage[colorlinks]{hyperref}
\hypersetup{%
    plainpages=true,
    breaklinks=true,
    hypertexnames=false,
    pageanchor=true,
    colorlinks=true,
    linkcolor={cyan},
    citecolor={cyan},
    urlcolor={cyan},
    anchorcolor={black}
}
\usepackage{gensymb}
\usepackage{hhline}
\usepackage{braket, siunitx}
\usepackage{ifthen}
\newboolean{showcomments}
\setboolean{showcomments}{true}

\usepackage{cleveref}
\usepackage{xcolor}
\usepackage{bbding} 
\usepackage{comment}
\usepackage{xspace}
\usepackage{hyperref}
\usepackage{tabularray}

\usepackage{chngcntr}

\crefname{equation}{Eq.}{Eqs.}
\Crefname{equation}{Equation}{Equations}
\crefname{figure}{Fig.}{Figs.} 
\Crefname{figure}{Figure}{Figures}
\crefname{section}{Sect.}{Sects.}
\Crefname{section}{Section}{Sections}
\crefname{table}{Table}{Tables}
\crefname{appsec}{}{Appendices} 

\newcommand{\mynote}[3]{%
  \ifthenelse{\boolean{showcomments}}{%
   \fbox{\bfseries\sffamily\scriptsize#1}%
   {\small$\blacktriangleright$\textsf{\emph{\color{#3}{#2}}}$\blacktriangleleft$}}%
  {%
   \@bsphack
   \@esphack
  }%
}

\newcommand{\panel}[1]{({#1})}

\newcommand*{\figref}[2]{%
  \hyperref[{#1}]{%
    ~\ref*{#1}%
    \ifx\\#2\\%
    \else
      \panel{#2}%
    \fi
  }%
}

\renewcommand{\selectlanguage}[1]{}
\begin{document}
\preprint{APS/123-QED}
\widetext

\title{Improving Transmon Qubit Performance with Fluorine-based Surface Treatments}

\def\LLaffil{Lincoln Laboratory, Massachusetts Institute of Technology, Lexington, MA 02421, USA}
\def\RLEaffil{Research Laboratory of Electronics, Massachusetts Institute of Technology, Cambridge, MA 02139, USA}
\def\Physaffil{Department of Physics, Massachusetts Institute of Technology, Cambridge, MA 02139, USA}
\def\EECSaffil{Department of Electrical Engineering and Computer Science, Massachusetts Institute of Technology, Cambridge, MA 02139, USA}

\author{Michael~A. Gingras}\email{michael.gingras@ll.mit.edu}\affiliation{\LLaffil}
\author{Bethany~M.~Niedzielski}\affiliation{\LLaffil}
\author{Kevin~A.~Grossklaus}\affiliation{\LLaffil}
\author{Duncan~Miller}\affiliation{\LLaffil}
\author{Felipe~Contipelli}\affiliation{\LLaffil}
\author{Kate~Azar}\affiliation{\LLaffil}
\author{Luke~D.~Burkhart}\affiliation{\LLaffil}
\author{Gregory~Calusine}\affiliation{\LLaffil}
\author{Daniel~Davis}\affiliation{\LLaffil}
\author{Renée~DePencier~Piñero}\affiliation{\LLaffil}
\author{Jeffrey~M.~Gertler}\affiliation{\LLaffil}
\author{Thomas~M.~Hazard}\affiliation{\LLaffil}
\author{Cyrus~F.~Hirjibehedin}\affiliation{\LLaffil}
\author{David~K.~Kim}\affiliation{\LLaffil}
\author{Jeffrey~M.~Knecht}\affiliation{\LLaffil}
\author{Alexander J. Melville}\affiliation{\LLaffil}
\author{Christopher~O'Connell}\affiliation{\LLaffil}
\author{Robert~A.~Rood}\affiliation{\LLaffil}
\author{Ali~Sabbah}\affiliation{\LLaffil}
\author{Hannah~Stickler}\affiliation{\LLaffil}
\author{Jonilyn~L.~Yoder}\affiliation{\LLaffil}
\author{William~D.~Oliver}\affiliation{\RLEaffil}\affiliation{\Physaffil}\affiliation{\EECSaffil}
\author{Mollie~E.~Schwartz}\affiliation{\LLaffil}
\author{Kyle~Serniak}\email{kyle.serniak@ll.mit.edu}\affiliation{\LLaffil}\affiliation{\RLEaffil}

\date{\today}

\begin{abstract}
Reducing materials and processing-induced decoherence is critical to the development of utility-scale quantum processors based on superconducting qubits.
Here we report on the impact of two fluorine-based wet etches, which we use to treat the silicon surface underneath the Josephson junctions (JJs) of fixed-frequency transmon qubits made with aluminum base metallization.
Using several materials analysis techniques, we demonstrate that these surface treatments can remove germanium residue introduced by our JJ fabrication with no other changes to the overall process flow.
These surface treatments result in significantly improved energy relaxation times for the highest performing process, with median $T_1=334~\mu$s, corresponding to quality factor $Q=6.6\times10^6$.
This result suggests that the metal-substrate interface directly underneath the JJs was a major contributor to microwave loss in these transmon qubit circuits prior to integration of these surface treatments.
Furthermore, this work illustrates how materials analysis can be used in conjunction with quantum device performance metrics to improve performance in superconducting qubits.
\end{abstract}

\maketitle
\section{Introduction}
Superconducting quantum circuits are widely regarded as a promising platform to realize utility-scale quantum computation~\cite{kjaergaard_superconducting_2020}.
High-fidelity single- and two-qubit gates have been demonstrated in a variety of Josephson-junction-based superconducting qubit architectures~\cite{sung_realization_2021, negirneac_high-fidelity_2021, ding_high-fidelity_2023, zhang_tunable_2024, li_realization_2024}, which are commonly limited by the coherence of the individual qubits.
To improve these coherence times, many efforts have focused on reducing material defects and those induced by fabrication processing~\cite{oliver_materials_2013,muller_towards_2019, siddiqi_engineering_2021} --- an important avenue of research that results in a performance advantage regardless of the specific superconducting qubit architecture~(e.g., the transmon~\cite{koch_charge-insensitive_2007} or fluxonium~\cite{ manucharyan_fluxonium_2009}).

\begin{figure}[t]
    \centering
    \includegraphics[scale=1]{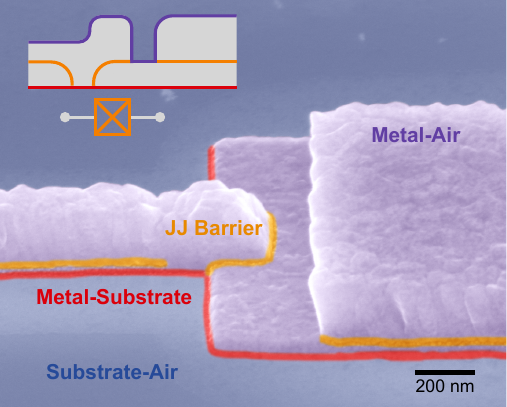}
    \caption{Dielectric interfaces that contribute to energy loss in superconducting quantum circuits, as illustrated in a colorized scanning electron micrograph of a Josephson junction (JJ), taken at a tilt angle of approximately $52\degree$ from normal to the surface. Inset is a cross-sectional cartoon of the multilayer JJ heterostructure, with the location of the intentional JJ indicated by the alignment of the JJ circuit element below. In this work, we apply pre-deposition surface treatments which modify the metal-substrate interface underneath the JJs of our qubits.}
    \label{fig:interfaces}
\end{figure}

A wide range of low-microwave-loss superconducting materials, including Al, Nb, Ta, NbTiN, TiN, and granular Al~\cite{bruno_reducing_2015,grunhaupt_loss_2018,place_new_2021,altoe_localization_2022,bal_systematic_2023, biznarova_mitigation_2024,ganjam_surpassing_2024}, can be patterned using conventional lithographic techniques to form the wiring and embedding circuitry necessary for superconducting qubit control, readout, and coupling.
However, with relatively few exceptions, the Josephson junctions (JJs) that are crucial to superconducting qubit construction are fabricated as superconducting tunnel junctions from Al/AlOx/Al heterostructures.
These are typically fabricated using double-angle shadow-evaporation lift off techniques~\cite{dolan_offset_1977,lecocq_junction_2011,bilmes_probing_2022,smirnov_wiring_2023}, which some regard as an obstacle to high-yield manufacturability.
Furthermore, the dielectric interfaces associated with JJ fabrication have been identified as hosts of lossy defects that can limit coherence and are often modeled as ``two-level systems" (TLS)~\cite{martinis_decoherence_2005,muller_towards_2019}.
Understanding and mitigating decoherence induced by interactions with these TLSs is an outstanding problem in the field today.
These TLS are understood to reside at various dielectric interfaces (see Fig.~\ref{fig:interfaces} and Refs.~\cite{wenner_surface_2011,wang_surface_2015,calusine_analysis_2018, woods_determining_2019}), including the metal-substrate interface underneath the JJs~\cite{dunsworth_characterization_2017}, the parasitic junctions associated with double-angle evaporation~\cite{bilmes_probing_2022}, and the JJ dielectric barrier itself~\cite{mamin_merged-element_2021}.
It has been shown that these interfaces in and around the JJs contribute strongly to decoherence, inspiring efforts to reduce defects in JJ fabrication and move toward more scalable, industrially compatible fabrication processes~\cite{verjauw_path_2022, van_damme_high-coherence_2024}.
Regardless of the techniques used to fabricate JJs, understanding and mitigating decoherence induced by this processing is critical to improving the performance of superconducting qubits.

In this article, we report on the improvement of transmon qubit performance by the use of fluorine-based surface treatments and argon ion milling --- prior to JJ deposition steps --- to reduce defects and residues at the metal-substrate interface around the JJs.
Over a dataset spanning nearly 200 qubits over 6 fabrication variations (all with Al base metallization on Si substrates), we identify a process combination that gives a statistically significant improvement in the quality factor of $2\times$ as compared to our process of record (POR).
Using several materials characterization techniques, we identify which processing steps reduce specific types of defects and contaminants, and attribute the aforementioned quality factor improvement to the removal of germanium-containing residues introduced by the processing of our shadow evaporation mask.
The highest performing process, which combines a commercially available Al ``pad etch" treatment and Ar ion milling before JJ deposition, yields a median energy relaxation time $T_1=334~\mu$s, corresponding to a median quality factor $Q=6.6\times10^6$ over 42 measured qubits.
Using a combination of cryogenic qubit characterization and materials analysis represents a tool with which to identify, understand, and mitigate dominant materials-induced sources of decoherence in superconducting qubits.

\begin{figure*}
    \centering
    \includegraphics[scale=1]{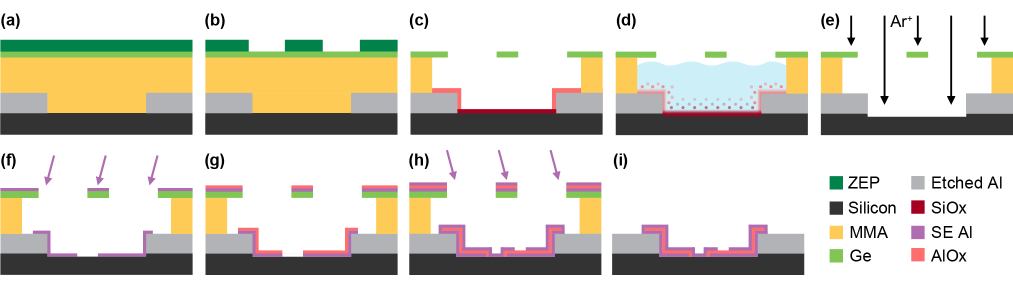}
    \caption{JJ fabrication process specific to this work. (a) A tri-layer resist stack of MMA, Ge, and ZEP is prepared on  patterned Al base metal. (b) Electron beam lithography creates a Dolan bridge pattern into the ZEP. (c) A $\mathrm{CF}_4$ plasma transfers the pattern into the Ge layer and an isotropic $\mathrm{O}_2$ plasma removes exposed MMA, growing oxide on the underlying Si and Al surfaces. (d) An optional F-based surface treatment (either 1\%~aqueous~HF or ``pad etch") removes byproduct residues and oxides grown during step (c). (e) \emph{In vacuo} Ar ion milling further treats the surface and opens connection to base metal Al immediately prior to JJ deposition. (f) A first angled deposition of Al forms the bottom JJ electrode. (g) Thermal oxidation of the Al surface forms the JJ tunnel barrier. (h) A JJ counter-electrode is deposited at an angle equal and opposite to that of step (f). (i) A liftoff process removes the sacrificial Dolan stack, completing the patterning of the JJ.}
    \label{fig:process}
\end{figure*}

\section{Qubit Fabrication}
 
In this section, we outline the fabrication processes for the fixed-frequency transmons included in this study.
Our experiment builds upon a POR fabrication flow, from which we study the effects of optional process modifications.
The main focus of this work is the effect of liquid, fluorine-based surface treatments prior to JJ deposition on substrate-metal interface losses relative to our baseline process.
Additionally, we report on the effects of strap connections (also called bandages or band-aids) between the JJ leads and base metallization~\cite{dunsworth_characterization_2017} and pre-JJ-deposition Ar ion milling.
The overall fabrication process flow is shown in Figure \ref{fig:process}.

In our baseline process flow, a double-side polished, un-doped, float-zone, high-resistivity~($>10~\mathrm{k}\Omega\cdot \mathrm{cm}$), prime-grade 50~mm (100) silicon wafer is cleaned with standard 50:50  $\mathrm{H}_2$$\mathrm{SO}_4$:$\mathrm{H}_2$$\mathrm{O}_2$ piranha solution for 15 minutes.
Wafers then get a 2~minute dip in a 1:49 HF:$\mathrm{H}_2$O acid solution to remove the native silicon oxide.
After this substrate preparation, we deposit a 250~nm high-quality Al base metallization layer at ultra-high vacuum in a molecular beam epitaxy (MBE) system.
This wafer is optically patterned using Merck AZ1512 photoresist and  wet-etched using Transene Aluminum Etchant Type A.
Wafers are stripped immediately after etch in room temperature acetone.
This defines the coarse features of the device design, such as the feedline, readout resonators, and majority of the qubit capacitor islands~(see Fig.~\ref{fig:single_qubit}a for partial optical micrograph of these features in the device under test).
AZ1512 series resists of thickness 1.5 $\mu$m and 3.0 $\mu$m are then used to create a simple, optical resist stack for definition of high-Z scanning electron beam lithography (SEBL) alignment marks.
A bilayer of 20~nm~Ti and 100~nm~Pt is evaporated onto this stack and lifted off with Microposit\textsuperscript{TM} Remover 1165 to create the alignment marks.

To pattern the JJs, a three-material stack of MMA/Ge/ZEP is used to create a Dolan-bridge style resist mask~\cite{dolan_offset_1977}.
Similar Ge-based resist stacks have been reported on before~\cite{nakamura_100-k_1996}, and our specific process described here has been used to demonstrate high-coherence qubits~\cite{yan_flux_2016, ding_high-fidelity_2023}.
First, a sacrificial layer of Kayaku copolymer MMA(8.5)MAA EL-11 is spin coated to a thickness of 450~nm and then vacuum-baked at 170\textdegree~C for 150~min to create a stable base for the hard mask.
A film of tensile-stressed Ge is then deposited in a high-vacuum electron beam evaporation system.
This is followed by a spin coating of electron-beam sensitive resist (ZEP520A from Zeon Specialty Materials).
Coarse and fine JJ mask features are exposed using a Raith EBPG 5200 electron-beam lithography system using beam currents of 100~nA and 1~nA, respectively, and the wafer is developed in Zeon ZED-N50, then cleaned with MIBK:IPA 1:3 and IPA.
After development, a reactive ion etch (RIE) system using $\mathrm{CF}_4$ gas etches the Ge mask.
A low power, isotropic $\mathrm{O}_2$ plasma on the same system is then used to clear and undercut the MMA.

At this point, step-height profilometry is performed on the remaining resist mask to determine the total thickness of the MMA/Ge stack.
This data is then used to optimize subsequent shadow deposition angles so as to achieve a specific JJ overlap regardless of upstream process variation.
Wafers are staged for shadow deposition and receive a high power $\mathrm{O}_2$ plasma treatment on the RIE system for an additional surface descum.
With this Dolan-style lithography mask exposing the underlying metal-substrate interface, optional surface treatments can be performed prior to JJ deposition.

Silicon oxide and surface residues may be generated during the plasma etching of the Ge/MMA stack.
Lossy oxides~\cite{quintana_characterization_2014, chayanun_characterization_2024} and etch byproducts should ideally be mitigated by reducing the use of the plasma steps in question, adjusting dry etch chemistry, or by removing the oxides prior to shadow evaporation.
To that end, in this work we compare two surface treatments prior to JJ deposition. 
First, a dilution of 1:100 HF:$\mathrm{H}_2$O was used to test efficacy of an acid readily available in most cleanrooms.
The wafers are submerged and agitated for 60 seconds.
While 1\% HF readily removes surface oxides, it also has a highly variable aluminum etch rate which can lead to undercutting of the exposed Al base metallization.
With this HF processing, straps were used to ensure connection between the junctions and the base metal, as we will discuss later.

To reduce etching of the underlying base metal, a commercially available pad etchant was also tested, specifically Silox Vapox III~\cite{dai_maskless_2006} from Transene.
This class of etchants is designed to open final passivation layers in semiconductor process flows for bond pad connections and have high selectivity to underlying metals.
The chemistry tested (henceforth referred to as ``pad etch") is a buffered acid etchant specifically designed for use in the presence of Al, though other pad etchants with like components could yield similar results.
This HF, ammonium fluoride, and acetic acid based wet etchant is infused with Al corrosion inhibitors and readily removes Si surface oxides.
Treatment with pad etch consisted of the wafer being submerged and agitated for 10 seconds.
This surface treatment resulted in only mild etching of the base metal Al, allowing for comparison of devices with either in-situ ion milled connections between JJs and the base metal or strap connections.

The wafer is then loaded into a Plassys MEB550SL3 electron-beam evaporator for JJ deposition.
This system includes a connected loadlock, ion mill chamber, deposition chamber, and oxidation chamber with automated fixture handling.
The junction deposition process has been described by Dolan~\cite{dolan_offset_1977} and is commonly used in the superconducting qubit field.
Figure \ref{fig:dolan_mask_JJ} in Appendix A shows a typical Ge Dolan style bridge after MMA etching along with a fully fabricated JJ made with a similar size bridge.
An optional \emph{in vacuo} ion mill can be performed prior to deposition to make contact between the junction and base metal aluminum.
This mill uses Ar as the working gas, an accelerator grid voltage of 175~V, and beam current of 50~mA.
Standard milling time is 75~s with a cryo-pumped chamber pressure of $\sim1\times10^{-4}$~Torr during the process.

Wafers are then immediately transferred into the high-vacuum evaporation chamber on the system and deposited with 30~nm of aluminum on an unheated wafer holder to create the first JJ electrode.
During deposition, the wafer is tilted at the angle determined by the aforementioned profilometry step to target a fixed JJ overlap dimension of 200~nm.
Static JJ oxidation is performed in a separate, dedicated chamber using pure semiconductor-grade oxygen up to a dose specific for the desired JJ critical current density ($J_c$).
Pressures can range from 10~mTorr for high $J_c$ devices up to 100~Torr for low $J_c$ devices.
All of the devices fabricated in this study targeted a $J_c\approx 1~\mu\mathrm{A}/\mu\mathrm{m}^2$.
With JJ lengths in the 100-150~nm range, and a fixed 200~nm overlap, this results in qubits with frequencies from 3-3.5~GHz.

After oxidation, the wafer returns to the evaporation chamber to deposit the 160~nm Al JJ counter-electrode at similar conditions to the first deposition, but deposited at opposite tilt angle.
Lastly, a controlled capping oxidation is performed prior to the wafer leaving the deposition system.
Junction materials are lifted off in vessels of acetone at 50\textdegree~C without sonication.

The pre-JJ Ar ion milling creates a robust connection between shadow evaporated JJ leads and the base metal. 
However, if the 1\% HF pre-JJ surface treatment is used, it completely removes the exposed base metal, so a straps process~\cite{dunsworth_characterization_2017} is used to subsequently make connection between the shadow-evaporated JJ electrodes and base metal layers.
Due to the extent of the base metal attack, large $14~\mu\mathrm{m}~\times~14~\mu\mathrm{m}$ straps were needed.
To determine any impact of these straps, devices with pad etch surface treatment were fabricated with and without straps of a more standard $6~\mu\mathrm{m}~\times~14~\mu\mathrm{m}$ size in addition to the larger size. 
Examples of both strap designs combined with 1\% HF and pad etch treated JJ overlap connections can be found in Figure \ref{fig:straps_fig} in Appendix B.

The straps are fabricated using a photolithography resist stack process similar to that used for SEBL alignment marks and patterned using standard techniques.
The wafer is then loaded back into the Plassys MEB550SL3 for Ar-ion milling \emph{in vacuo} using similar conditions as the pre-JJ deposition ion milling.
300~nm of Al is then electron-beam evaporated into a region overlapping both the base metal and shadow evaporant.
This deposition is done at an angle normal to the wafer, under similar conditions to the JJ evaporation, and is followed by lift off in heated acetone.

As a final step of fabrication, we deposit photolithography-defined Al air-bridges to suppress parasitic slotline modes throughout the chip~\cite{rosenberg_3d_2017,rosenberg_solid-state_2020}.
Completed wafers are diced into chips with a protective resist coating using a diamond dicing saw.
The coating is removed with an acetone and IPA clean prior to packaging for measurement.

\section{Qubit Measurements}
\begin{figure}
    \includegraphics[scale=1]{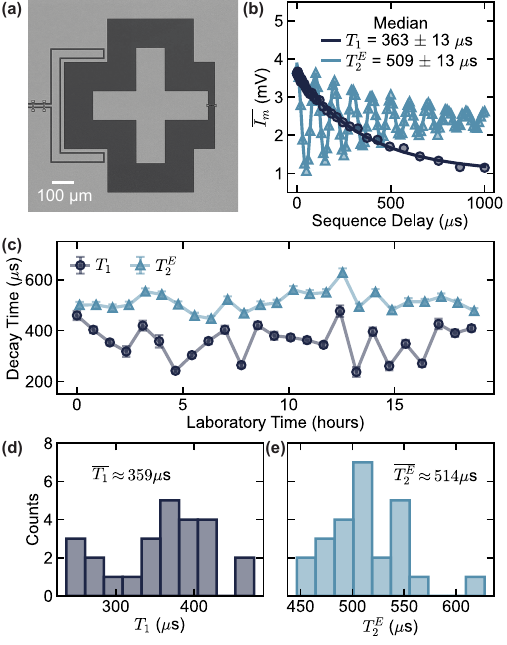}
    \caption{Characterization of coherence times of a qubit fabricated using Process D: pad etch treatment, pre-JJ deposition Ar ion milling, and straps, shown as an example. (a) Optical micrograph of a representative transmon qubit, nominally identical to those characterized in this study. The gap between single-ended capacitor island and ground plane is~$130~\mu$m on all sides except for the where the JJ resides, which is $30~\mu$m. (b) $T_1$ and $T_2^E$ decay curves corresponding to the median measured values for the qubit, where $\overline{I_m}$ is the average in-phase quadrature component of the demodulated dispersive readout signal. (c) Fluctuations of the measured decay timescales over time, acquired over the course of $\sim 19$ hours for this particular qubit. (d) and (e) depict histograms of the measured decay timescales reported in (b).}
    \label{fig:single_qubit}
\end{figure}
All chips under test were patterned with six uncoupled, single-ended transmon qubits with relatively large island to ground gap to reduce loss associated with the metal-air and metal-substrate interfaces of the Al base metallization~\cite{geerlings_improving_2012} (see~Fig.~\ref{fig:single_qubit}a).
Each qubit is accompanied by a dedicated readout resonator coupled to a common feedline in a hanger geometry.
The chips were wire bonded into two-port microwave packages machined from 6061 Al in which signals were routed to the chip via Au-patterned alumina microwave interposers.
These were thermally anchored to the mixing chamber stage of a cryogen-free dilution refrigerator with a base temperature of 10--15~mK.
All readout and qubit control signals were applied through the common feedline using standard microwave components and techniques.
Individual chips were characterized in a single cooldown, though our combined dataset was acquired over multiple cooldowns between four functionally equivalent dilution refrigerators and microwave test setups.

\begin{table*}[ht!]
\centering
\begin{tblr}{Q[l,m] Q[l,m] Q[l,m] Q[l,m] Q[c,m] Q[c,m] Q[c,m] Q[c,m] Q[c,m] Q[c,m]} \hline\hline
Process & Treatment & Ar Mill & Straps &  $Q\ (10^6)$ & $ T_1$ ($\mu$s)  & $T_2^E$ ($\mu$s)  & Q CV (\%) & {Qubits\\Measured}\\[2.5pt]
\hline
POR & None & Yes & No & $3.3\ ^{+0.4}_{-0.8}$ &  $157\ ^{+24}_{-40}$  &  $190\ ^{+54}_{-27}$  & 32.1 & 59/60\\[2.5pt]
A & HF & No & Yes & $4.2\ ^{+2.8}_{-0.8}$  & $209\ ^{+126}_{-44}$  &  - & 54.1 & 9/12\\[2.5pt]
B & HF & Yes & Yes & $5.0\ ^{+0.7}_{-1.0}$ &  $232\ ^{+54}_{-43}$  &  $283\ ^{+2}_{-17}$  & 31.5 & 16/18\\[2.5pt]
C & Pad Etch & No & Yes & $5.1\ ^{+1.6}_{-1.7}$ &  $283\ ^{+94}_{-103}$  &  $199\ ^{+32}_{-38}$  & 39.0 & 28/30\\[2.5pt]
D & Pad Etch & Yes & Yes & $5.3\ ^{+2.8}_{-2.4}$ &  $262\ ^{+97}_{-130}$  & $254\ ^{+67}_{-105}$ & 46.0 & 28/36\\[2.5pt]
E & Pad Etch & Yes & No  & $6.6\ ^{+1.6}_{-1.5}$ &  $334\ ^{+89}_{-86}$   & $271\ ^{+53}_{-39}$ & 30.7 & 42/48\\[2.5pt]
\hline\hline
    \end{tblr}
\caption{Summary statistics for six qubit process conditions. Values represent process medians +/- 25\% quartile. We quantify ``Qubits Measured" as the number of qubits included in our $T_1$ dataset relative to the total number cooled down for testing (for more details on this definition, see the text of Section~\ref{sec:results}.)
}
\label{table:perm}
\end{table*}

The energy relaxation time $T_1$ and the single-echo decoherence time $T_2^E$ of each qubit were all characterized using standard microwave pulse-probe techniques~\cite{krantz_quantum_2019}.
An example coherence dataset from a representative qubit treated with pad-etch, Ar ion milling, and straps deposition is shown in Fig.~\ref{fig:single_qubit}.
Each of these decay timescales presented were characterized using ``free-decay" techniques in which the qubit state is allowed to evolve under the influence of its environment during a variable time denoted ``sequence delay" (in between intentional control pulses that differ for each experiment), before we perform a dispersive measurement to project the state of the qubit.
Notably for echo-decoherence experiments (see Fig.~\ref{fig:single_qubit}, we utilize a sequence-delay-dependent virtual $Z$ rotation to the second $\pi/2$ pulse in order to generate oscillations in the decay signal, which results in an exponentially-decaying cosine functional form that enables more reliable extraction of $T_2^E$ than an exponential decay alone.

Each decay timescale is sampled sequentially in a round-robin style at least 25 times over the course of 5-20 hours, depending on the repetition rate chosen for each individual experiment (see Fig.~\ref{fig:single_qubit}b) and the degree to which measurements across different qubits were multiplexed in time.
In most instances, measurements of the six qubits on a given chip were interleaved unless one was miscalibrated, in which case it was typically (but not always) re-measured separately after correction.
The average of those measurements is reported as representative of the qubit.
An example of the characterization performed on each qubit is shown in Fig.~\ref{fig:single_qubit}.
In addition, Ramsey decoherence experiments were performed on all qubits, in which we sometimes observed of ``beating" which can be indicative of strong coupling to two-level system defects, and in some of these instances we were unable to reliably extract decay timescales. 
We chose to not report Ramsey decoherence results in this manuscript in order to avoid biasing these coherence times toward those not effected by beating (which may plausibly skew larger).

\section{Results}
\subsection{Qubit Performance}\label{sec:results}
\begin{figure}[hb!]
    \centering
    \includegraphics[scale=1]{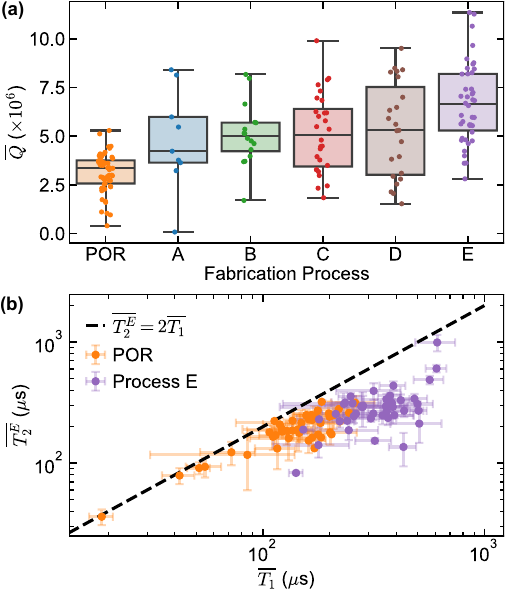}
    \caption{Summary of coherence times across all surface treatments, ion mill conditions and straps configurations tested in this study. (a)  Box and whisker plot with points jittered illustrating the median (center horizontal line), inter-quartile range (box), and total range (whiskers) of measured qubit quality factors $\overline{Q}=\omega_q\times \overline{T_1}$ for each process shown in Table~\ref{table:perm} (b) Qubit-average Hahn-echo coherence time $\overline{T_2^E}$ vs average energy relaxation time $\overline{T_1}$ for the baseline POR process and the highest performing process tested, Process E. Error bars represent one standard deviation of the measured values for each qubit.}
    \label{fig:perm}
\end{figure}

Qubits from wafers with combinations of surface treatment, pre-JJ ion mill, and straps were fabricated and then characterized at cryogenic temperatures.
Qubit quality factor $Q$ was chosen as the metric by which to compare qubits within and between different processes: $Q = \omega_q T_1$ where $\omega_q$ is the qubit transition frequency in angular units.
Our analysis builds off the premise that our qubits are limited by the underlying average material loss tangents (supported by our results) and we assume that these, and therefore this $Q$ metric, are frequency independent.
Each data point in Figure~\ref{fig:perm} corresponds to the average Q for a given qubit measured as described in the previous section, amounting to 182 qubits characterized for this study.

The median summary statistics in Table~\ref{table:perm} and boxplot data in Figure \ref{fig:perm} compare our baseline process, which has only pre-JJ Ar ion milling, against all process splits with surface treatment before junction deposition and various combinations of ion milling and straps.
This data shows a statistically significant performance improvement over our baseline POR process in all the surface-treated splits~(see Table~\ref{table:pvalue}).
Additionally, all splits with straps processing show similar performance despite differences in ion milling, contrary to expectation from literature~\cite{dunsworth_characterization_2017}. 
The median Q is slightly higher for ion-milled vs non-milled splits of the same surface treatment when straps are included, but not with statistical significance~\ref{table:pvalue}.
The highest-performing process split received pad etch surface treatment and pre-JJ ion milling with no straps processing, in which we found a median~$Q=6.6\times10^6$ and~$T_1=334~\mu$s which we compare to~$Q=3.3\times10^6$ and~$T_1=157~\mu$s for our POR process.
This $2\times$~improvement over POR with these process conditions is statistically significant~\ref{table:pvalue} and this process split is also statistically improved over all other conditions.
Of note is the significant improvement of the pad etch, mill, no straps condition over the similar condition with straps, in which we measure median~$Q=5.3\times10^6$ and~$T_1=262~\mu$s.

We note that not all qubits that were cooled down were included in our $T_1$ dataset, which for completeness is reported in Table~\ref{table:perm} as ``Qubits Measured."
We choose this definition to encompass failures in our automated calibration that were not corrected (and therefore excluded from our analysis), in addition to instances of fabrication yield loss, which is difficult to define quantitatively.
For example, as each chip is designed to support frequency-multiplexed measurements of six qubits, driven through a common feedline, intentional variation of qubit frequencies is necessary.
The $J_c$ realized on some wafers was lower than targeted, resulting in some inconveniently low qubit frequencies, either in terms of readout contrast or introduction of significant charge dispersion that affects both qubit dephasing~\cite{schreier_suppressing_2008} and readout~\cite{serniak_direct_2019}.
We note that this was more common on wafers with pre-JJ deposition surface treatments, which could indicate an additional influence of these treatments on $J_c$ targeting.
This can be corrected by updates to our JJ oxidation parameters, and we have no evidence that this is correlated with the coherence improvements reported on in this article.
Future work will explore differences in $J_c$ targeting between our POR and surface treated qubits, motivated by reports that processing of the sub-JJ surface can impact the rate of JJ aging~\cite{pop_fabrication_2012}.

Finally, we note that not all qubits in the ``Qubits Measured'' category were included in our reporting of process median $T_2^E$, as some qubits showed evidence of qubit frequency or other miscalibration that would not appreciably impact the extracted $T_1$, but could influence $T_2^E$~(as an example, Stark shifts could be induced at small pulse durations~\cite{chen_measuring_2016}).
We do not expect that these exclusions would be related to any of the fabrication processes probed here, nor should they systematically bias the reported~$T_2^E$ coherence times.

We observe that across all process splits, the individual qubit average $\overline{T_2^E}$ is typically short of the energy relaxation limit imposed by $2\overline{T_1}$~[Fig.~\ref{fig:perm}b].
Future work will aim to identify this dominant dephasing channel, which could be explained by photon-shot-noise dephasing from residual occupation of the readout resonators, dynamics of dielectric TLS population or frequency distribution occurring at fast timescales.

\subsection{Materials Characterization}
To determine the cause of the improvement in measured qubit $Q$ with surface treatments, JJs deposited with four different pre-deposition treatments were characterized by cross-sectional scanning transmission electron microscopy (STEM).
The four STEM samples were of a JJ with no treatment, with and without ion milling, and with the pad etch treatment, with and without ion milling.
These samples were prepared by $\mathrm{Ga}^+$ focused ion beam (FIB), with cross-sections taken longitudinally through the center of the junctions.
Bright-field (BF) and annular dark-field (ADF) STEM imaging was carried out in an aberration-corrected STEM operating at 200~kV.
STEM samples were additionally examined with energy dispersive x-ray spectroscopy (EDS) and electron energy loss spectroscopy (EELS) to determine the elemental distribution within the samples.

\begin{figure}
    \centering
    \includegraphics[scale=1]{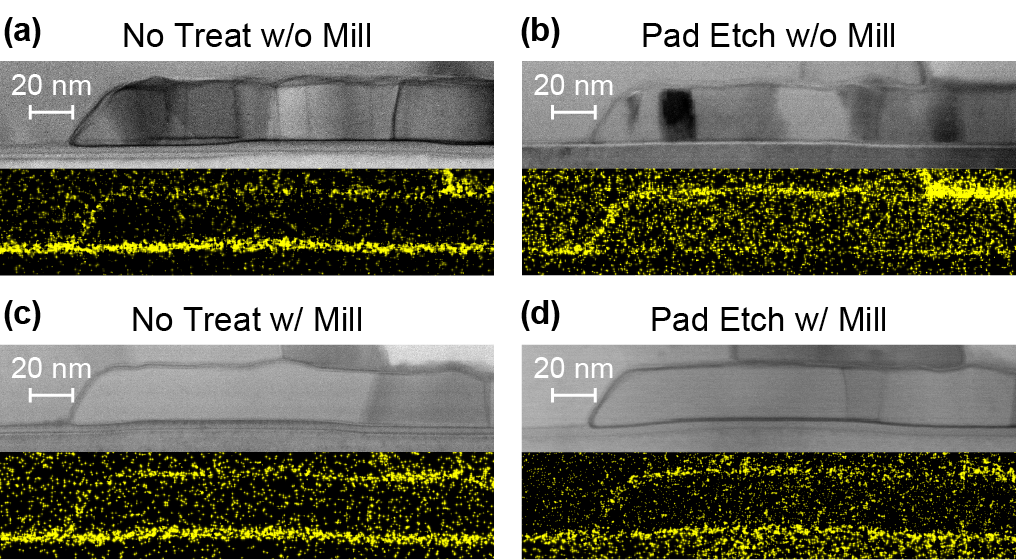}
    \caption{BF STEM images (top) and EDS Oxygen maps (bottom) of cross-sections taken from JJs that received four different pre-deposition processes as indicated above each image. a) Shows a strong silicon oxide present under the JJ without treatment(s). b) Shows the oxide effectively removed by pad etching, leaving just a native oxide. c) Shows POR milling alone is not effective at removing oxide. d) Shows that even with pad etch, milling regrows a detectable oxide.}
    \label{fig:tem}
\end{figure}

BF STEM images of each junction cross-section and corresponding EDS oxygen maps are shown in Figure \ref{fig:tem} revealing a surprising interplay between the surface treatment and ion milling.
Both junctions with no pad etch treatment  (Figure \ref{fig:tem}a and c) show an oxide present underneath the junction at the metal substrate interface, as indicated by the EDS oxygen map.
This oxide can be removed using a pad etch (Figure \ref{fig:tem}b), but is present again after the sample receives ion milling prior to junction deposition (Figure \ref{fig:tem}d).
This suggests that the ion mill rebuilds an oxide at the metal-substrate interface.
This result was unexpected, as the ion mill step prior to the first aluminum evaporation is completed with inert argon and the wafer is under high vacuum for the duration of the time before aluminum evaporation.
It is also interesting that this process split showed better qubit performance despite containing this sub-JJ oxide layer.

\begin{figure}
    \centering
    \includegraphics[scale=0.38]{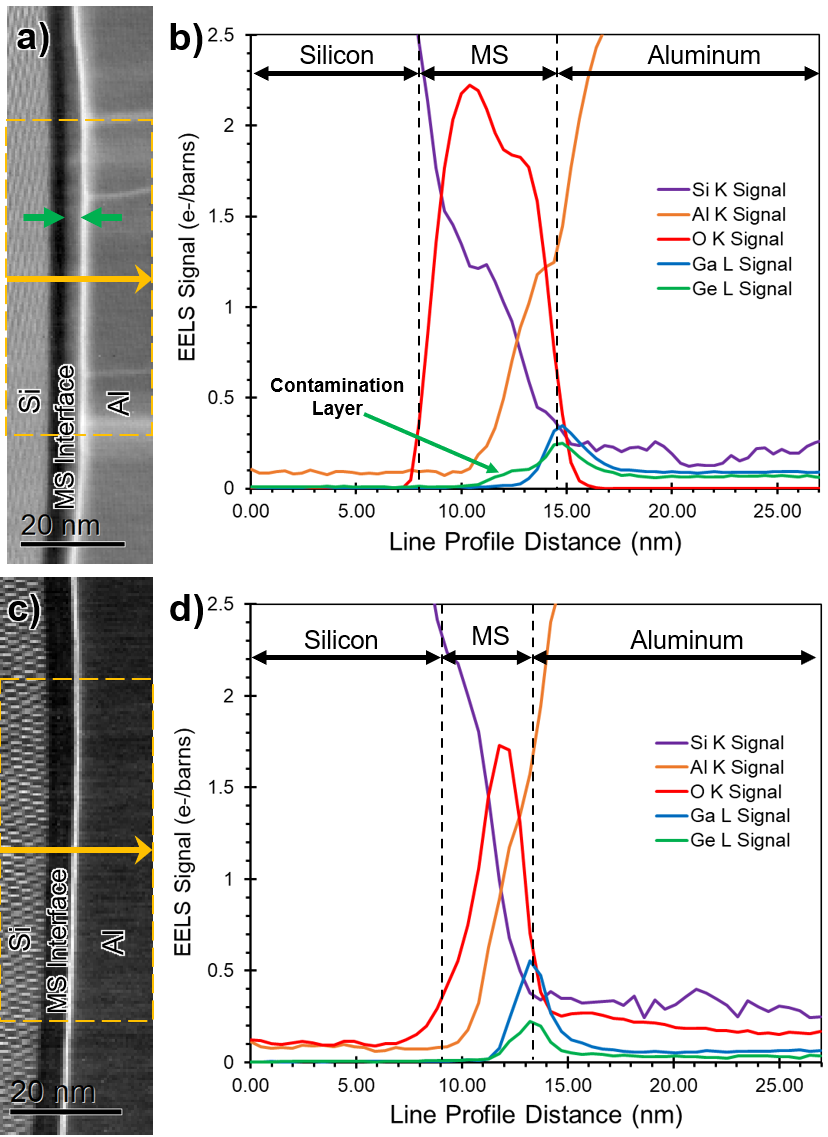}
    \caption{ADF STEM images and EELS signal profiles from the metal-substrate interface region of two pre-treatment process conditions. a) shows an ADF image from the sample that received no pad etch or ion mill before junction deposition, and b) shows an EELS elemental signal profile extracted from mapping of the area shown in a) and taken in the direction indicated by the orange arrow and vertically averaged over the area indicated by the orange box in a).  Similarly c) shows an ADF image of the sample that received both a pad etch and ion mill before junction deposition, and d) Shows an EELS signal profile in the direction and from the area indicated by the orange box and arrow in c).  A contamination layer is visible under the Al layer in a) as indicated by green arrows.  The Ge L edge signal profile of the corresponding location in b) indicates presence of Ge in the contamination layer.}
    \label{fig:eels}
\end{figure}

This effect was probed further using higher resolution STEM imaging and EELS mapping of the metal-substrate interface under the junction.
This metrology was done to compare a sample with neither surface treatment nor pre-JJ ion milling, and a sample with both a pad etch surface treatment and ion milling.
ADF STEM imaging of the metal-substrate interface from the sample that received no pad etch or ion mill reveals a possible contamination layer in the oxide layer below the aluminum, visible as the area of brighter contrast and indicated by green arrows in Figure \ref{fig:eels}a.
Imaging of the sample that received a pad etch and ion mill (Figure \ref{fig:eels}c) does not show a similar layer.
EELS elemental signal profiles across the metal-substrate interface area were extracted from EELS spectrum mapping of those same areas.
The location, direction, and area averaged to generate those profiles are indicated by the yellow dotted boxes and arrows in Figures \ref{fig:eels}a and \ref{fig:eels}c.
In the EELS signal profile for the sample that received no surface treatment (Figure \ref{fig:eels}b), the Ge L edge signal indicates the presence of Ge in the possible contamination layer.
In contrast, the Ge signal is absent from the oxide layer in the EELS profile for the sample that received both a pad etch and ion mill (Figure \ref{fig:eels}d).
This absence indicates that the combined action of the pad etch and ion mill is able to remove a Ge containing contamination layer that would otherwise be present.
Ga contamination from FIB sample preparation was present in both mapped areas under the Al/interface boundary and along grain boundaries.
This results in a false Ge signal in these areas due to overlap in the location of the EELS Ga L edge and the following Ge L edge.
The Ge contamination indicated in Figure \ref{fig:eels}b may be concluded to be from the presence of Ge only because no corresponding Ga signal is present at that position.

\begin{figure}[hbt!]
    \centering
    \includegraphics[scale=1]{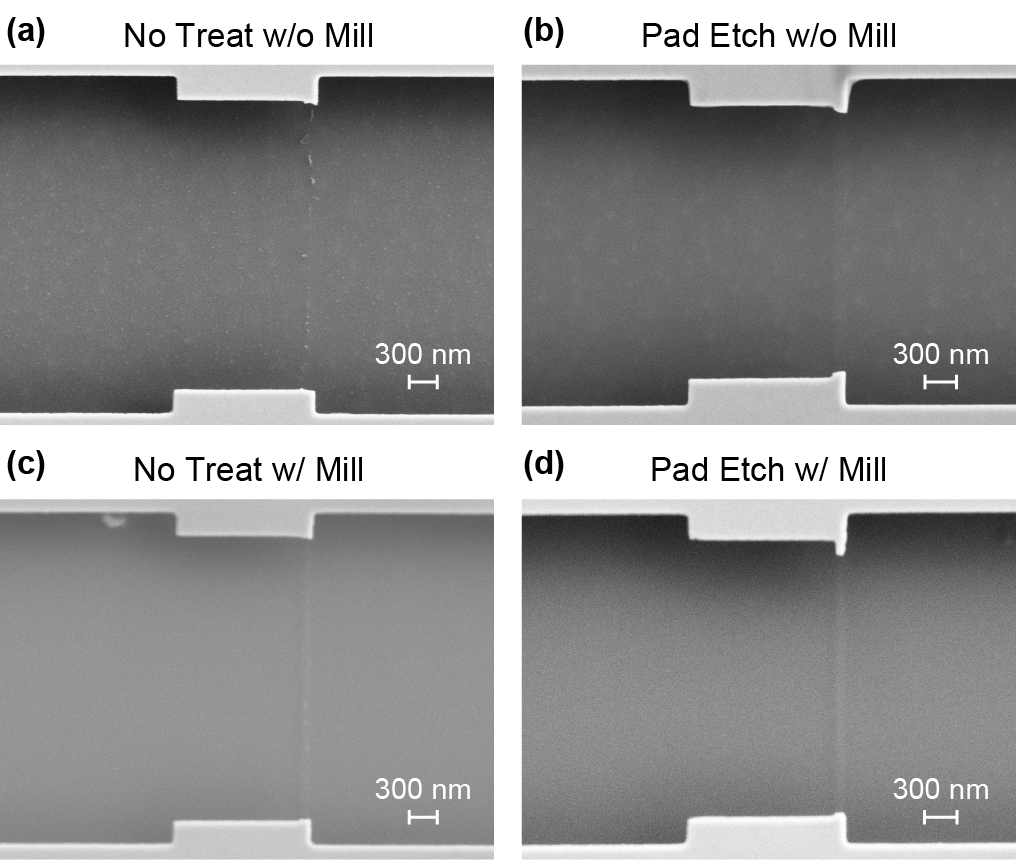}
    \caption{Top-down SEM images of JJ shadow masks with Dolan bridges removed. a) Shows the native surface prior to treatment - residue is widespread across the trench area and collected under the removed bridge. b) Shows a sample with pad etch and without pre-JJ ion milling - residue under bridge has been removed but the trench surface still shows contamination. c) Shows a sample without pad etch and with pre-JJ ion milling - trench residue is eliminated but small amounts still exist under the removed bridge. d) Shows a sample that received both pad etch and pre-JJ ion milling - both areas of the sample are clean. 
}
    \label{fig:residue}
\end{figure}

To determine the mechanism behind the efficacy of the pad etch and mill, four wafers were prepared through Ge/MMA dry etch and up to the point of pad etch processing.
Each sample then received the process combinations labeled above the scanning electron microscope (SEM) images in Figure \ref{fig:residue}, a sample with no treatment, with and without ion milling, and with the pad etch treatment, with and without ion milling.
The germanium Dolan bridges shown in Figure \ref{fig:dolan_mask_top_down} in Appendix A  were removed to check for the contamination that was observed in the EELS data and high resolution STEM imaging.
The untreated samples both showed a thin line of residue present where the bridge had been, while pad etched samples were clean in this area.
The samples without ion mill both showed a faint residue on the Si away from where the bridge existed, while the milled samples were significantly cleaned in this region.
The pad etched and ion milled sample was the cleanest of the four, with no residues present in either location.
Samples with either mill or pad etch both showed faint outlines where the Dolan bridge existed, suggesting a shadowing of the silicon surface during the hard mask dry etching, the ion mill, or both.

Based on the EELS of contamination and SEM of residues, the likely source of these residues is germanium dry etch byproducts being carried to the silicon surface during the subsequent MMA oxygen dry etch.
These residues accumulate underneath the bridge where only surface treatments can remove them.
The residues on the Si, away from the bridge, were too thin to be characterized, but could potentially be germanium-based and deposited via a similar mechanism to those under the bridge.
Removal of this residue under the Dolan bridge is shown to likely be the most vital to boosting qubit quality, due to its proximity to where the qubit electric field is highest.
The ion milling, which both removes residues away from the junction and re-grows an oxide at the substrate metal interface, is less of a factor.

\section{Conclusions and Outlook}

In summary, the use of surface treatments prior to JJ deposition effectively improved the substrate-metal interface immediately below the Dolan bridge in our transmon qubits.
We attribute a statistically significant increase in overall qubit quality to the removal of germanium-based residues.
When using a pad etch designed for use with aluminum base metal and a pre-deposition ion mill, the improvement was enhanced and led to a $2\times$ increase in qubit quality.
The source of this improvement was investigated using SEM, TEM, EDS and EELS characterization, narrowing the potential root cause to hard mask dry etch products accumulating under the Dolan bridge.
These byproducts are blocked from the ion mill prior to deposition, but can be removed with surface treatment.
Despite the ion mill re-growing an oxide layer underneath the JJ, the removal of Si surface abnormalities away from the JJ could explain the enhanced qubit performance with this process combination.
In addition to the benefit from surface cleans, a significant improvement from removing strap connections on treated qubits is seen --- this increase in performance warrants further investigation to see how strap processing limits our qubits once the metal substrate interface is cleaned.

These results highlight the benefit of an iterative characterization and performance evaluation process that helped diagnose and mitigate a major source of loss in our qubits.
Identifying this source also reinforces the benefit of improving the metal-surface interface using surface cleans prior to base metal and JJ depositions~\cite{yano_fabrication_1987, melville_comparison_2020, altoe_localization_2022, chayanun_characterization_2024, kopas_enhanced_2024, bland_2d_2025, colao_zanuz_mitigating_2025}.
Future work will continue to use this approach and focus on further exploring new etch conditions to reduce interfacial oxides, residues and substrate damage during Dolan mask patterning.
In parallel, alternative electron beam resist stack materials, both organic and inorganic~\cite{savu_quick_2009,welander_shadow_2012,tsioutsios_free-standing_2020}, will be explored to further reduce etch byproducts and allow for JJ deposition under more optimal conditions, including higher temperatures and ultra-high vacuum environments.

\section{Acknowledgments}
We gratefully acknowledge M. Augeri, P. Baldo, R. Das, K. Harmon, M. Hellstrom, A. Kurlej, C. Reed, M. Ricci, C. Thoummaraj, and D. Volfson at MIT Lincoln Laboratory for technical assistance. 
This research was funded under Air Force Contract No. FA8702-15-D-0001. 
Any opinions, findings, conclusions or recommendations expressed in this material are those of the authors and do not necessarily reflect the views of the US Air Force or the US Government.

\appendix
\counterwithin{figure}{section}
\counterwithin{table}{section}
\label{Appendix}

\section{Ge Dolan Bridge Masks}

\begin{figure}[h!]
    \centering
    \includegraphics[scale=0.95]{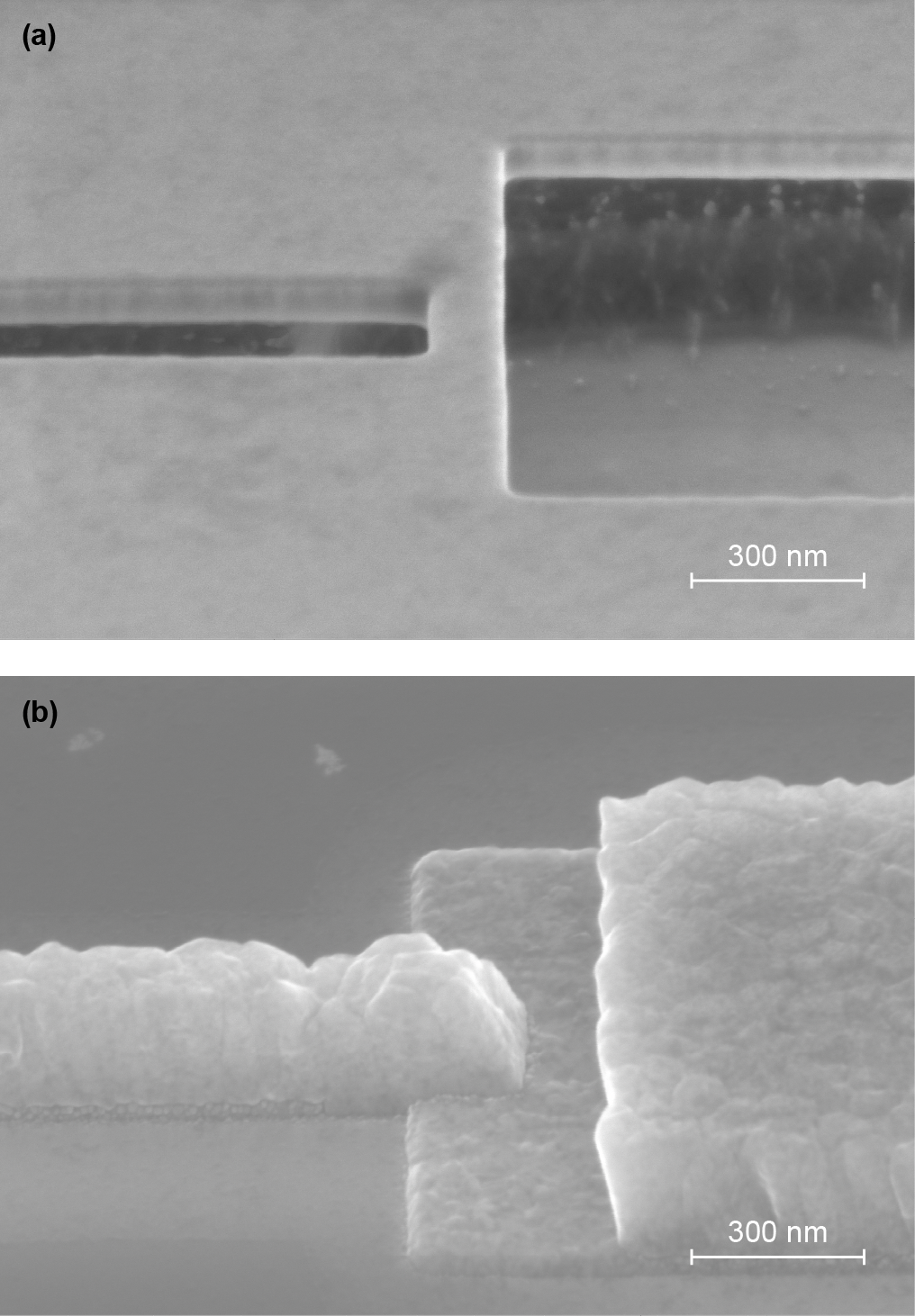}
    \caption{SEM images of a Ge Dolan bridge mask and JJ. (a) Shows the Dolan bridge shadow mask used to make JJs prior to deposition. (b) Shows a completed JJ post liftoff of the Ge and MMA layers.}
    \label{fig:dolan_mask_JJ}
\end{figure}

Figure \ref{fig:dolan_mask_JJ} shows a set of tilted SEM images of a typical Ge/MMA Dolan shadow bridge before and after JJ fabrication.
In image a) the top layer is 100~nm of Ge, the bottom layer is 450~nm of undercut MMA.
In b) the JJ shown is of nominal size 200x200~nm, yielding a barrier area of $\approx 0.04~\mu\mathrm{m}^2$.

\begin{figure}[h!]
    \centering
    \includegraphics[scale=0.98]{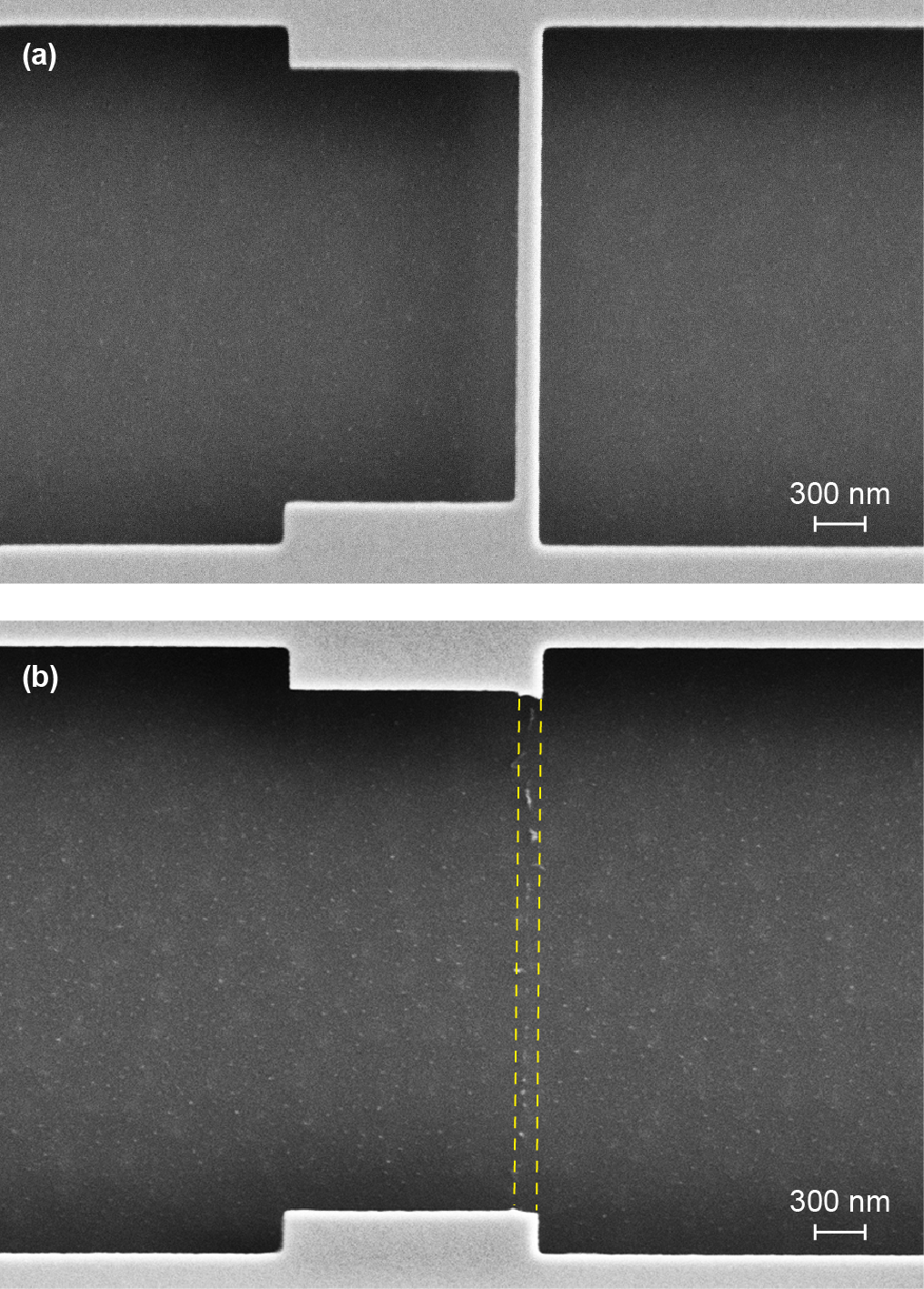}
    \caption{SEM images looking top-down at a Ge Dolan bridge mask, with the areas of lighter contrast indicating Ge. (a) shows the Ge bridge as fabricated and (b) shows the bridge removed with dashed lines to guide the eye to its original location. The darker areas show the exposed substrate surface, which have not yet received treatment or ion mill.}
    \label{fig:dolan_mask_top_down}
\end{figure}

Figure \ref{fig:dolan_mask_top_down} shows a top down SEM image of a longer, nominal size 2500x200~nm size Dolan bridge prior to JJ fabrication.
This figure gives an example of a bridge prior to removal for residue inspection.

\section{JJ Strap Connections}

\begin{figure*}[h!]
    \centering
    \includegraphics[scale=1.5]{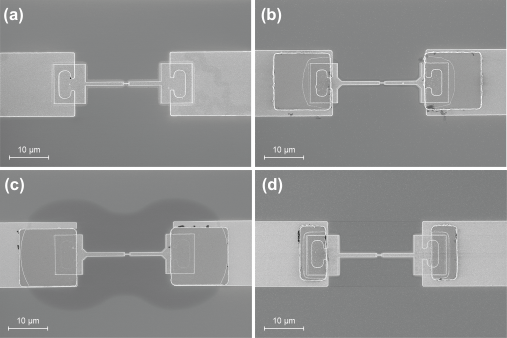}
    \caption{A series of four SEMs showing the shadow evaporated JJ leads and their overlap connection to the aluminum base metallization. Each image shows a different combination of JJ pre-treatment and strap connection. a) No pre-JJ wet treatment, with Ar ion mill and no straps (POR) b) Pad etch treated surface, with no Ar ion mill and ``large strap" c) LHF treated surface, no Ar ion milling, and ``large" strap. d) Pad etch treated JJ, with Ar ion mill, and ``standard" strap}
    \label{fig:straps_fig}
\end{figure*}

Figure \ref{fig:straps_fig} shows a series of top down SEM images of the shadow evaporated JJ leads and their overlap connections to the base metallization interconnects. These micrographs illustrate the effects of different pre-treatments and show the geometry of our different strap designs.

\section{Qubit Statistical Test Data}
\label{fig:ttest}

Table \ref{table:pvalue} shows calculated p-values for a matrix of comparisons on all process combinations tested in this work.
Using the Tukey-Kramer HSD all pairs means test method, p-values are calculated for each process comparison.
P-values of less than 0.05 represent a rejection of the null hypothesis of equal means, meaning the average qualities of each process combination are statistically different to a 95\% confidence.

\begin{table*}[h!]
\centering
\begin{tblr}{Q[l,m] | Q[l,m] | Q[l,m] Q[l,m] Q[l,m] Q[l,m] Q[l,m] Q[l,m]} \hline\hline   

& N & POR  & A & B & C & D & E\\
\hline
POR & 59 & 1 & 0.0130 & 0.0002 & $<$.0001 & $<$.0001 &  $<$.0001 \\
A & 9 & 0.0130 & 1 & 0.6743 & 0.5802 & 0.4380 & 0.0052\\
B & 16 & 0.0002 & 0.6743 & 1 & 0.9063 & 0.6895 & 0.0037 \\
C & 28 & $<$.0001 & 0.5802 & 0.9063 & 1 & 0.7405 &  0.0009 \\
D & 28 & $<$.0001 & 0.4380  & 0.6895 & 0.7405 & 1 & 0.0052 \\
E & 42 & $<$.0001 & 0.0052 & 0.0037 & 0.0009 & 0.0052 & 1 \\
\hline\hline

\end{tblr}
\caption{Qubit quality statistical test results (P-values) with each process combination of treatment statistically compared. P-values of less than 0.05 represent processes that are statistically different to a 95\% confidence.}
\label{table:pvalue}
\end{table*}

\clearpage
\newpage 

\bibliography{references.bib}
\end{document}